ChatGPT & Mechanical Engineering: Examining performance on the FE Mechanical Engineering and Undergraduate Exams

Authors

Prof. Matthew E Frenkel
matthew.frenkel@nyu.edu
(646) 997-3518
Bern Dibner Library
5 MetroTech Center, Room 303
Brooklyn, NY 11201
ORCID: 0000-0002-6883-1105

Hebah Emara ()
he2156@nyu.edu
(646) 997-3527
Bern Dibner Library
5 MetroTech Center, Room 302
Brooklyn, NY 11201
ORCID: 0009-0006-9642-2225

Abstract

The launch of ChatGPT at the end of 2022 generated large interest into possible applications of artificial intelligence in STEM education and among STEM professions. As a result many questions surrounding the capabilities of generative AI tools inside and outside of the classroom have been raised and are starting to be explored. This study examines the capabilities of ChatGPT within the discipline of mechanical engineering. It aims to examine use cases and pitfalls of such a technology in the classroom and professional settings. ChatGPT was presented with a set of questions from junior and senior level mechanical engineering exams provided at a large private university, as well as a set of practice questions for the Fundamentals of Engineering Exam (FE) in Mechanical Engineering. The responses of two ChatGPT models, one free to use and one paid subscription, were analyzed. The paper found that the subscription model (GPT-4) greatly outperformed the free version (GPT-3.5), achieving 76% correct vs 51% correct, but the limitation of text only input on both models makes neither likely to pass the FE exam. The results confirm findings in the literature with regards to types of errors and pitfalls made by ChatGPT. It was found that due to its inconsistency and a tendency to confidently produce incorrect answers the tool is best suited for users with expert knowledge.

Introduction

The introduction of the electronic calculator revolutionized the jobs of engineers. The tedious and time-consuming task of consulting trigonometric and logarithmic tables was reduced to pressing a few buttons. In much the same way, the calculator also changed the engineering classroom for students and teachers. The electronic calculator is just one of many disruptive technologies introduced in the 20th and 21st centuries that altered the landscape of the engineering workplace and classroom. Computer Aided Design (CAD) software changed the process of engineering drafting. Personal and high performance computing changed the timescales needed for complex calculations and simulation. The modern internet remade the platforms used for communication and collaboration. Once the utility of these disruptive technologies is understood by the engineering workforce, their adoption becomes ubiquitous and as a result they also become an important element of engineering education. University students around the world are entering classrooms with calculators and laptops in hand, collaborating on learning management systems, and engineering software such as MATLAB, Ansys, and SOLIDWORKS are being directly integrated into classroom curriculum.

Recent developments in publicly available artificial intelligence technologies, specifically large language models (LLMs) may be the next disruptive technology that will reshape the engineering profession and engineering education. Generative AI, made possible through LLMs, such as ChatGPT have seen tremendous growth in usage recently and the profession has begun to react. Academic journals are implementing policies around ChatGPT authorship, universities around its usage inside and outside the classroom, companies are determining how to integrate it into workflows, and in some cases countries are deciding whether to restrict or ban access (Stokel-Walker, 2023; Grose, 2023; Satariano, 2023). Since its release researchers have tested ChatGPT's capabilities in a variety of fields. In this paper, we look to expand on the work of these researchers by testing ChatGPT's ability to answer mechanical engineering questions. Our research intends to address the following questions: how does ChatGPT perform in the field of mechanical engineering, and what use cases may exist for it in the classroom or in professional settings?  To this end first, we will share a literature review that provides important background information.  Next, we will explain how we determined what questions to ask ChatGPT and how we collected our data. Following the methodology, we will present the results of this study, highlighting the difference between the freely available ChatGPT model and the subscription GPT-4 model. Finally, we will conclude the paper with a discussion of the results, context we believe is important for integrating ChatGPT into engineering education, and future directions of research we plan to pursue.

## Literature review

### ChatGPT Background

Within the first week of its launch on November 30, 2022, ChatGPT, developed by OpenAI (2022), reached 1 million users, making it the fastest growing application to date (Dennean et al., 2023). ChatGPT (Generative Pre-Trained Transformer) is a language model capable of understanding text and producing human-like responses (Hashana et al., 2023). LLMs like ChatGPT are trained on massive text datasets and aim to predict the next word in a sequence (Kung et al., 2023 as cited in Hashana et al., 2023). The Transformer architecture supports this process by identifying patterns in how words, phrases, and sentences relate to each other and serves as the foundation for ChatGPT's ability to generate text (Nield, 2023; Hashana et al., 2023). In contrast to supervised language learning models, OpenAI's use of semi-supervised learning has allowed ChatGPT to leverage large quantities of unlabeled diverse texts, such as Wikipedia articles, in its unsupervised pre-training phase without the need for the additional data processing and labeling required for supervised learning tasks such as classification (Wu et al., 2023; Delua, 2021; Radford et al., 2018). After pre-training, supervised learning is then used to fine-tune the model based on specific natural language understanding tasks (Radford et al., 2018). Between the first iteration of GPT in June 2018 and the release of GPT-3 in May 2020, the model's parameters and pre-training data have grown significantly from 117 million parameters and about 5 GB of pre-training data to 175 billion parameters and 45 TB of pre-training data (Wu et al., 2023). GPT-4 was released in March 2023 as a large multimodal model that is capable of accepting both image and text inputs; however, image input is not currently available for public use (OpenAI, 2023c). Alongside the release of GPT-4, OpenAI's release notes for ChatGPT indicate that it is far from a static tool and show that the company releases updates and adds functionality to ChatGPT roughly every two weeks (Natalie, 2023). OpenAI also launched a ChatGPT app for iOS devices in May 2023 and has announced plans to incorporate third party plugins, such as the technical computing Wolfram plugin, with ChatGPT as well as an experimental model that "knows when and how to browse the internet" (OpenAI, 2023e; OpenAI 2023d, para. 18).

**ChatGPT & Exams**

OpenAI and researchers from various disciplines have already begun testing ChatGPT's performance on standardized exams and its capacity to solve complex problems. This review will focus on ChatGPT's performance on STEM-related exams and datasets. While methodologies, the version of ChatGPT used (i.e., GPT-4, GPT-3, ChatGPT, etc.), and the decision to use prompt engineering to improve performance vary among these studies, the results provide valuable insight on ChatGPT's strengths and weaknesses.

Regarding medical exams, Humar et al. (2023) found that while ChatGPT did not pass the US Plastic Surgery In-Service Exam, it had "an impressive baseline medical knowledge" but was unable to answer questions that required complex and context-dependent decision making (p.3). Fijačko et al (2023) similarly found that while ChatGPT was able to provide useful explanations in its answers, it failed to pass the American Heart Association Basic Life Support and Advanced Cardiovascular Life Support exams. Kasai et al. (2023) gave both GPT-4 and ChatGPT questions from the Japanese Medical Licensing Exam in Japanese and found that while GPT-4 outperformed ChatGPT and passed the exam, it underperformed compared to medical student averages and selected "prohibited" answers that would harm patients, adding to the notion that ChatGPT and GPT-4 struggle to make complex decisions. Glison et al. (2023) found that while ChatGPT was able to pass the US Medical Licensing Exam, achieving a score equivalent to a third year medical student, its accuracy was significantly lower for more difficult questions. Fergus et al. (2023) further supported that ChatGPT struggles to solve questions that require complex analysis and application of knowledge in a pharmaceutical sciences setting.

In their analysis of ChatGPT's capabilities, Laskar et al. (2023) found that ChatGPT performs best in social science and worst in STEM and suggested that its performance in math revealed its tendency to answer questions correctly but with faulty reasoning or with the correct reasoning but the wrong answer. Shakarian et al. (2023) also identified ChatGPT's weakness in math, finding that as the complexity of a math word problem increases (ex., higher numbers of additions and subtractions required), so does

ChatGPT's failure rate. This study also found that there was a significant improvement in performance (~60%) when ChatGPT was asked to show its work, showing that the use of prompt engineering or modifying prompts can produce clearer answers from ChatGPT (Shakarian et al., 2023). Plevris et al. (2023) further emphasized that while ChatGPT was able to provide convincing answers, it did not produce reliable or consistent results on complex mathematical problems or advanced logical tasks and often provided long and detailed incorrect answers without acknowledging any possible errors. Azaria et al. (2023) also stated that "ChatGPT demonstrates overconfidence in the mathematical field that surpasses its actual capabilities" (p.12). Xuan-Quy et al.'s (2023) results also found that ChatGPT confidently provided false information and statements on a national high school physics exam in Vietnam and stated that ChatGPT struggled to answer questions that require a deep understanding of physics. Yuan et al.'s (2023) analysis found that while ChatGPT was capable of answering arithmetic questions correctly in some cases it often provided estimations that are close to the correct answer, revealing it did not actually calculate the mathematical answer. In their research on how LLM behavior changes over time, Chen et al. (2023) found that between the March 2023 and June 2023 versions, GPT-4 performed significantly worse (from 97.6% to 2.4% accuracy) when asked to identify whether a number was prime while GPT-3.5 performed significantly better (from 7.4% to 86.8% accuracy).

Frieder et al.'s (2023) broad analysis of ChatGPT's mathematical capabilities further supported the model's limitations, stating that it performed significantly below the average mathematics graduate student, answering questions inconsistently and confidently providing incorrect answers. In their assessment of ChatGPT's ability to solve real world problems in oil and gas industrial engineering, Ogundare et al. (2023) found that ChatGPT faltered when tasked with applying knowledge to novel or unusual scenarios. Giannos and Delardas' (2023) assessment of ChatGPT's performance on UK standardized admissions exams also found that ChatGPT underperformed in science and math and when asked questions that require content-specific or prior knowledge of a given topic. Pursnani et al. (2023) found that while GPT-4 was able to achieve about 75% accuracy on questions from the Environmental Fundamentals of Engineering (FE) exam, it failed when answering complex multi-step calculations. When the correct formula was selected, GPT-4 still arrived at the wrong answer (Pursnani et al., 2023). Koubaa et al.'s (2023) analysis of ChatGPT's performance on challenges pulled from the IEEExtreme computer programming competition emphasized that "ChatGPT still has limitations in understanding and solving complex programming problems that require high-level reasoning and expertise" (p.7).

In their evaluation of GPT-4 (both with and without visual input), OpenAI (2023c) highlighted that GPT-4 outperformed its predecessor, GPT-3.5, and achieved high passing scores on STEM standardized exams such as AP Biology, AP Chemistry, AP Calculus BC, AP Physics 2, and the SAT and GRE math and quantitative sections. While GPT-4 outperformed its predecessor and other large language models, the model was still prone to several limitations that include: hallucinating (fabricating information), confidently providing wrong answers, lack of knowledge past September 2021, an inability to learn from experience, and "various biases" (OpenAI 2023a, p.11). Bordt and

von Luxburg (2023) stressed that while GPT-4 outperformed GPT-3.5 on an undergraduate computer science exam, its ability to pass the exam "[did] not imply that it has any understanding of computer science" (p.6). Borji (2023) corroborated these sentiments, stating that models like ChatGPT lacked the "capability to reason about the connections between concepts and entities [and] can only generate text based on the patterns they have learned during training" (p.3).

These studies indicated that while ChatGPT was capable of passing several exams or performing to various degrees of competency on a variety of datasets, it was not a perfect tool. As a language model, ChatGPT did not appear to actually perform mathematical calculations, but instead predicted the answer in the same manner it predicts other text responses (Kortemeyer, 2023). Combining this with ChatGPT's verbose and confident responses, inconsistency, and its tendency to hallucinate results in misleading or wrong answers that appear correct adds to its limitations when answering complex STEM questions. Additionally, ChatGPT struggled with applying theory and knowledge to unique or novel situations which further supported the notion that it had no real understanding of concepts (Ogundare et al., 2023; Frieder et al., 2023).

**ChatGPT & Engineering Education**

ChatGPT has already had a disruptive impact on education and scholarly research, with several universities and schools debating how to address potential student use and researchers attempting to list ChatGPT as a co-author on scholarly research articles (Hartman-Sigal, 2023; Stokel-Walker, 2023; van Dis et al., 2023). This section will focus on ChatGPT's performance as a "student" in STEM courses and will outline instances where ChatGPT can be used as a supplemental tool to help students and instructors and instances where it may cause more harm than good.

Deshpande and Szefer (2023) assessed ChatGPT's performance in an introductory computer engineering course and found that while it performed well on questions related to general concepts, its inability to process or produce images and difficulty with hands-on labs made it incapable of passing the course. Kortemeyer (2023) found that ChatGPT was able to narrowly pass an introductory calculus based physics course but emphasized its inconsistency and frequent numerical errors when performing calculations. While ChatGPT achieved a B- in an introductory functional programming course, Geng et al. (2023) emphasized that without prompt engineering the AI performed significantly worse and highlighted that even for students who opted to use ChatGPT to complete their homework, there was no significant improvement in their homework grades. Elder et al. (2023) similarly found that while ChatGPT passed their sophomore level design course with a 73% grade, it performed best with follow up questions and additional explanations for more complex problems.

Increasing dialogue among engineering educators, developing best practices and guidelines for generative AI use in the classroom, and establishing an emphasis on training for using these tools can lead to positive applications of ChatGPT in engineering education (Johri et al., 2023). ChatGPT may be a useful resource for developing syllabi, creating lesson plans, improving assignments, providing students with personalized feedback, and creating virtual learning environments - with the caveat that its responses can be inconsistent and inaccurate (Azaria et al., 2023; Geng et al., 2023; Menekse, 2023). It may also be possible to use ChatGPT to create tailored problem sets to challenge students at various proficiency levels (Azaria et al., 2023). ChatGPT can also be a useful test prep resource capable of providing test takers with more meaningful and detailed feedback than traditional test prep materials (Fijačko et al., 2023). As ChatGPT has been shown to be prone to biases and hallucinations, researchers have suggested employing more openness and transparency into the sources of ChatGPT's training data, ChatGPT's design, and how it works can help students and educators understand its limitations and how to effectively use ChatGPT (Wu et al., 2023; Borji, 2023; Hashana et al., 2023).

ChatGPT's negative impacts on engineering education may be more subtle than the concerns educators have on its potential to weaken students' writing skills and its impact on plagiarism. Chegg, a website that provides homework help, and other similar online resources have existed long before ChatGPT's launch and have already served as tools for students to find answers to homework and other problems, dulling the impact of ChatGPT as an outlet for cheating (Kortemeyer, 2023). ChatGPT's current inability to process or generate images also serves as a major limitation to its use in engineering and related subjects (this may change if GPT-4's image processing capabilities are made available through OpenAI's paid ChatGPT Plus service or its free version of ChatGPT) (Deshpande & Szefer, 2023; OpenAI, 2023b). There are, however, concerns that overreliance on ChatGPT might weaken students' critical thinking and problem solving skills and may lead to the passive acceptance of ChatGPT's answers as facts (Geng et al., 2023). ChatGPT has been inconsistent when providing references and would not be suited for beginners in a given subject as they may be misled into believing ChatGPT's incorrect but overly confident and heavily detailed answers (Plevris et al., 2023; Azaria et al., 2023; Sallam et al., 2023; Kortemeyer, 2023). Language models trained on human text are also imbued with the language humans use with regard to specific topics. ChatGPT, GPT-3, and GPT-4 have been shown to harbor negative attitudes towards math and physics, may exacerbate existing math anxiety in students, and may imply that certain groups are better at math than others, adding to the existing stereotype threat these students may already experience (Abramski et al., 2023; Maloney et al., 2013). Particular care appears necessary when using ChatGPT in the engineering classroom to address its biases and limitations, especially if presented as a tool for beginning learners.

Methods

The objective of this paper is to determine how successful ChatGPT is as a tool for answering engineering questions to better understand how it and other AI tools may impact engineering education in the future. To this end we searched for a set of questions that could be representative of one or more fields of engineering, that had unambiguous answers, and that were representative of expected knowledge of an undergraduate engineering student. The fundamentals of engineering exam (FE) produced by the National Council of Examiners for Engineering and Surveying (NCEES)

"is generally [a] first step in the process of becoming a licensed professional engineer" (NCEES, n.d., para.1). The test is 6 hours long and consists of 110 questions and is intended for college engineering students nearing graduation or recent graduates. There are versions of the test for a number of different engineering disciplines including but not limited to chemical, mechanical, and environmental engineering. The prominence of the exam also means numerous practice exams and review books have been created which provide a large set of potential questions. For these reasons it was determined that using questions from an FE practice exam would be the best way to move forward.

Using the FE exam was not without limitations. First, there are the numerous subject exams to decide between. As an objective of this paper we intended to analyze the responses produced by ChatGPT beyond the simple correct or incorrect to better understand any limitations of the tool that might exist. With this in mind it was determined that we should start with the mechanical engineering subject exam as one of the authors received their PhD in the field and currently teaches courses at the junior and senior level on the subject at a large private university. Second, many engineering questions are accompanied by diagrams, figures, tables, and other visuals that need to be viewed and understood to answer the question. At the time that this study took place, neither the free or paid versions of ChatGPT that were studied had the capacity to include an image input. With this in mind we determined that we would limit ourselves to only using questions that could be completely represented through text. Last, the copyright of the practice exam does not allow the text of questions to be redistributed in a publication such as this article. As a result we can only share the aggregated data collected from the FE exam questions and discuss the trends that were seen. In order to also provide examples of questions and responses we also included additional questions that were developed by one of our authors for use in their Automatic Control Theory and Heat Transfer course exams.

Recognizing these limitations, the authors developed a strategy for inputting questions to ChatGPT, collecting responses, and analyzing the results. The first step was to acquire a set of questions from an FE practice exam. The acquired set consisted of 100 questions with accompanying solutions. The questions were then analyzed to determine which could not be completely represented through text only input. This dropped the number to 47 questions consisting of multiple choice (39), fill in the blank (5), matching (1), and all that apply (2) formats. With the set of askable questions determined, each question was typed into a spreadsheet and categorized by both subject and question type. Questions that contained complex mathematical formulas were typed using LaTeX syntax. There are 14 different subjects included in the mechanical engineering FE exam. Table one provides these subjects, the number of corresponding questions that could be formatted for text input, the expected number of questions per subject expected on the FE exam, and the percent represented based on the lowest number of expected questions. The correct answer to each question was also included in the spreadsheet.

Table 1. Question subjects for the Mechanical Engineering FE exam and corresponding number of questions available and asked

| Analysis by question subject | Questions with text only input | Expected number on FE Exam | % Represented |
|---|---|---|---|
| Mathematics | 4 | 6-9 | 67% |
| Probability and Statistics | 4 | 4-6 | 100% |
| Ethics and Professional Practice | 4 | 4-6 | 100% |
| Engineering Economics | 3 | 4-6 | 75% |
| Electricity and Magnetism | 3 | 5-8 | 60% |
| Statics | 0 | 9-14 | 0% |
| Dynamics, Kinematics, and Vibrations | 2 | 10-15 | 20% |
| Mechanics of Materials | 1 | 9-14 | 11% |
| Material Properties and Processing | 6 | 7-11 | 86% |
| Fluid Mechanics | 4 | 10-15 | 40% |
| Thermodynamics | 6 | 10-15 | 60% |
| Heat Transfer | 2 | 7-11 | 29% |
| Measurements, Instrumentation, and Controls | 4 | 5-8 | 80% |
| Mechanical Design and Analysis | 4 | 10-15 | 40% |

It is worth noting that efforts were made to determine if visual inputs could be incorporated into the questions. At the time of the study there was no way to place an image directly into the chat box, but it was found that if a URL associated with an image on the open web was sent to ChatGPT it could be asked to describe that image. This feature was very inconsistent. At times the descriptions produced by chatGPT did not match the image at all. In cases when ChatGPT did recognize the general concept of an image, such as an electric circuit diagram, it still regularly misinterpreted the data in the image. For example, when shown an electrical circuit it claimed resistors in series were in parallel and took several additional prompts to correct this error. We also found that ChatGPT was unable to read images from URLs hosted in a Google Drive or Box regardless of the sharing settings. This reinforced the limitation of this study that only questions that could be formulated through text would be used.

Once all the questions were ready for text input, we began the process of collecting data from ChatGPT. Two different accounts were used for inputting the questions to ChatGPT. One account was restricted to the freely available version (GPT-3.5) of ChatGPT while the second account used GPT-4 via a paid subscription. In both cases all questions were asked individually, each question was asked as a "new chat". This allowed us to keep all answers separate and allowed for potential follow up after analyzing the response. Once a question was asked, the amount of time it took ChatGPT to complete its response was recorded. The full text of ChatGPT's response was then recorded and it was indicated on the spreadsheet if the response was correct or incorrect. Once all the questions were asked to both models their answers were further analyzed by the authors. This further analysis was done in an effort to better understand what happened when an answer was wrong, and to examine the explanations associated with correct answers. Following this analysis, we generated some follow up questions for GPT-4 to determine if follow up would improve its accuracy. The same process was followed for 13 questions developed by the author for their Automatic Control Theory and Heat Transfer Courses.

## Results/Discussion

### Sample Questions

As part of the protocol used in this paper a set of 13 questions used in exams for a 300 level mechanical engineering course in Automatic Control Theory and a 400 level course in Heat Transfer were given to both the freely available version of ChatGPT as well as the subscription based GPT-4. Overall GPT-4 answered 9 of 13 questions correctly and the freely available version answered 8 correctly. This can be further broken down as 6 questions that both models got correct, 2 questions that both models got wrong, 3 questions that GPT-4 got correct while the other model was incorrect, and 2 questions that the freely available version got correct while GPT-4 was incorrect. In this paper's appendix we provide Table A which contains the text of 4 sample questions as well as the responses given by both models. The questions and responses are presented exactly as they were given and received. In appendix A we have selected 1 question that both models got correct, 1 that both models got wrong, 1 that GPT-4 got correct while the free version was incorrect, and 1 with the opposite result. Additionally, the last row of Table A provides a free response question. In this case GPT-4 provides a correct answer while the free version does not. This additional question was included in the table to demonstrate the response to questions types other than multiple choice.

A closer examination of the responses provided in Table A demonstrates the positives and negatives offered by ChatGPT with respect to engineering questions. In the case where both provided the correct answer a significant similarity can be seen throughout both answers, but GPT-4's answer provides additional detail compared to the freely available GPT-3.5. Though these additional details do not impact the correctness of the response, they do provide more insight into the concept being tested by the question. This indicates that GPT-4 may be a better tool for a student attempting to better understand engineering concepts.

Looking at the sample question that GPT-4 got correct, but the free version did not, the answers again are very similar. Both GPT models start providing the correct equation necessary to complete the calculations and identify the correct information given the in question prompt. Moreover, both models are able to correctly setup the equation which needs to be solved, but while GPT-4 correctly solves the equation and converts to the proper units that matches the multiple choice option, the free version incorrectly converts from meters to micrometers, instead converting from meters to nanometers and improperly labels its result as micrometers. As a result the free version provides the incorrect answer. Unit conversion is a common mistake for an undergraduate student, but not the kind of error we anticipated seeing from ChatGPT. Unit errors like this showed up multiple times with regards to both the sample questions and the FE exam questions. It is important to recognize that both GPT models were guilty of this type of error at some point. From the perspective of a student learning mechanical engineering or a professional using ChatGPT in some capacity, this kind of error in a response could be easily overlooked due to how confidently GPT presented its response.

Moving next to the sample question that GPT-4 got incorrect while the free version was correct, there are again several similarities between both answers. It could even be argued that GPT-4 is as correct as the free version, but since it did not select one of the multiple choice answers it was marked as incorrect - as it would have been had it left the answer blank of a standardized test. The two models give slightly different definitions of the Biot number that is required of this problem. GPT-4 gives a more general equation in terms of a characteristic length and then substitutes that value with the radius. The free version skips the characteristic length step and provides the Biot number equation directly in terms of the radius. Both models correctly calculate a Biot number based on the equation they found and the numbers given in the problem. Both models also correctly indicate that their calculation satisfies the necessary condition for the Lumped Capacitance model which is the objective of this question. The reason that neither model produces a value that matches the choices provided is because both used the radius as the characteristic length. In class students were taught to determine the characteristic length by dividing the system's volume by its surface area. In the case of a sphere this would result in the radius divided by 3. The method used by the GPT models is a more conservative method and would not be considered wrong if this was a free response question instead of a multiple choice. After completing the calculation, GPT-4 recognizes its answer does not match any of the choices provided and includes at the end of its response a note that there could be an error in the question. On the other hand, the free model chooses the correct answer which is the only answer that both has a number that satisfies the lumped capacitance model and says so, but the model offers no explanation for the difference between its calculated value and the one in the selected answer.

In the sample question where both models were incorrect we again see great similarity between the answers. Both models correctly identify the physical equations needed to solve this problem and set up the equations correctly. In fact, each model gives the same equation for finding the coefficient of heat transfer:

$$h = \epsilon\sigma\frac{(T_1^4 - T_2^4)}{T_1 - T_2} = 0.8 * 5.67 * 10^{-8}\frac{(573.15^4 - 303.15^4)}{573.15 - 303.15}$$

Solving this formula should give the answer, *h=16.7*, but both models provide incorrect answers. GPT-4 answered with 10.2 and the free version with 1.36. This is a very interesting result as executing arithmetic is the last place a new user might expect to see an error from a software program, but is consistent with findings in the literature. This is likely a result of GPT being a predictive algorithm and not a calculator. This is important as people may assume that a software like ChatGPT would have no problem executing mathematical expressions and thus may not check its work, which can result in sharing incorrect information.

Last, we take a closer look at the free response question in the last row of Table A. This question has multiple parts and on an exam would have offered students an opportunity to earn partial credit by demonstrating an understanding of the material even if there were mistakes in calculations. Overall, GPT-4 provides a much better answer for this question. It completes part A correctly, gives reasonable properties for the materials in the problem, correctly calculates the Reynolds numbers and identifies the flow correctly as laminar. GPT-4 also uses the correct Nusselt number correlation for this flow type and ultimately calculates the heat transfer value to within a reasonable rounding error. On the other hand, the free model provides a less accurate set of material properties. It uses these properties to properly calculate the Reynolds number but it does not identify the flow regime properly. In fact, it states the problem does not specify the flow regime even though the Reynolds number has been determined which would have cost a student all the points for that section of the question. In the end it also fails in executing the arithmetic of its final expression for heat transfer. This problem demonstrates what was seen throughout all of the questions supplied to ChatGPT: that GPT-4 almost always provided a better answer with more context than the freely available model.

FE exam

An objective of this research was to explore the accuracy of the responses provided by the different models of ChatGPT. Our results showed that GPT-4 answered 36 of 47 questions correctly (76.6%) in a total time of 1853±188 seconds, approximately 30 minutes. The freely available version of ChatGPT answered 24 of 47 questions correctly (51.1%) in a total time of 1465±188 seconds, approximately 24 minutes. The FE exam allows 6 hours for completion and even accounting for the fact that we asked less than half the total questions on the exam it is clear that both models are capable of meeting the allowed timeframe. It is not possible to see if either model would pass the exam as the requirements for passing are not available, but it is clear that the accuracy of the models is substantially different, with GPT-4 clearly outperforming the freely available version.

In order to look deeper into the results we also analyzed the accuracy of responses by subject and by question type as seen in Tables 2, 3, 4, and 5. It is important to recognize the relationship between table 1 and tables 2 and 3 that focus on answers by question subject. Table 1 demonstrates that a number of subjects are underrepresented in this study. This is primarily due to the fact that some subjects are more prone to visual questions that require charts or diagrams and as a result could not be transformed into the text based input required of this study. 6 subject areas have less than 50% of their lowest expected number of questions represented: Statics, Dynamics, Mechanics of Materials, Fluid Mechanics, Heat Transfer, and Mechanical Design and Analysis. 3 subjects have between 50%-74%: Mathematics, Electricity and Magnetism, and Thermodynamics. The remaining 5 subjects are represented at 75% or greater: Engineering Economics, Probability and Statistics, Ethics and Professional Practice, Material Properties and Processing, and Measurement, Instrumentation, and Control.

Table 2.  Analysis of accuracy based on question subject for GPT-4

| Analysis by Question Subject: GPT-4 | Total Questions per Subject | Right by Subject | Percentage |
|---|---|---|---|
| Mathematics | 4 | 4 | 100.00% |
| Probability and Statistics | 4 | 4 | 100.00% |
| Ethics and Professional Practice | 4 | 4 | 100.00% |
| Engineering Economics | 3 | 2 | 66.67% |
| Electricity and Magnetism | 3 | 3 | 100.00% |
| Statics | 0 | 0 | Undefined |
| Dynamics, Kinematics, and Vibrations | 2 | 2 | 100.00% |
| Mechanics of Materials | 1 | 0 | 0.00% |
| Material Properties and Processing | 6 | 3 | 50.00% |
| Fluid Mechanics | 4 | 3 | 75.00% |
| Thermodynamics | 6 | 3 | 50.00% |
| Heat Transfer | 2 | 2 | 100.00% |
| Measurements, Instrumentation , and Controls | 4 | 3 | 75.00% |
| Mechanical Design and Analysis | 4 | 3 | 75.00% |

Table 3. Analysis of accuracy based on question subject for free version of ChatGPT

| Analysis by Question Subject: Free Version | Total Questions per Category | Right by category | Percentage |
|---|---:|---:|---:|
| Mathematics | 4 | 2 | 50.00% |
| Probability and Statistics | 4 | 3 | 75.00% |
| Ethics and Professional Practice | 4 | 2 | 50.00% |
| Engineering Economics | 3 | 0 | 0.00% |
| Electricity and Magnetism | 3 | 2 | 66.67% |
| Statics | 0 | 0 | Undefined |
| Dynamics, Kinematics, and Vibrations | 2 | 0 | 0.00% |
| Mechanics of Materials | 1 | 1 | 100.00% |
| Material Properties and Processing | 6 | 3 | 50.00% |
| Fluid Mechanics | 4 | 2 | 50.00% |
| Thermodynamics | 6 | 3 | 50.00% |
| Heat Transfer | 2 | 1 | 50.00% |
| Measurements, Instrumentation , and Controls | 4 | 3 | 75.00% |
| Mechanical Design and Analysis | 4 | 2 | 50.00% |

Comparing Tables 2 and 3 it is seen that GPT-4 performed as well as or better than the free version in nearly every subject. The only exception was Mechanics of Materials, but as this subject is only represented by a single question in the dataset it would be inappropriate to draw larger conclusions from it. Additionally, a closer examination of the response shows that both versions of ChatGPT calculated an incorrect value for this answer, but the free version chose the correct multiple choice option stating:

> None of the answer choices provided match this value exactly. However, the closest option is:
>
> C. 1,320
>
> Please note that the given answer choices might contain typographical errors or approximations, as the calculated value does not precisely match any of them.

More interestingly in this case C was not actually the closest answer to the value ChatGPT produced (2646.75) in the multiple choice options, but based on our methods since this was the stated answer it was marked as correct.

Overall, both versions of ChatGPT demonstrated that they are capable of answering engineering questions. GPT-4 provided accurate answers more often. GPT-4 had its worst performance, outside of the single Mechanics of Materials question, in the Thermodynamics and Material Properties and Processing. Both these sections were well represented with 6 questions, but in both subjects GPT-4 only answered 3 correctly. For many of the wrong answers in these subjects GPT-4 provided the correct methodology but required a material property or access to a chart or table that would have been included in the FE exam supplemental materials for an exam taker but was not available in the context of this study. In most cases GPT-4 stated that it would need more information to provide the calculations.

Tables 4 and 5 demonstrate that both models are capable of answering multiple choice, fill in the blank, matching, and all that apply questions. Both models struggled the most with fill in the blank questions (however, our dataset is small). This reflects that when calculations were off both models would sometimes still select a multiple choice answer that was closest to their calculation, but could not do this for fill in the blank type questions.

Table 4. Analysis of accuracy based on question type for GPT-4

| Analysis by Question Type: GPT-4 | Total Questions per type | Right by category | Percentage |
|---|---:|---:|---:|
| Multiple Choice | 39 | 30 | 76.92% |
| Fill in the Blank | 5 | 3 | 60.00% |
| Matching | 1 | 1 | 100.00% |
| Ordering | 0 | 0 | Undefined |
| All that apply | 2 | 2 | 100.00% |

Table 5. Analysis of accuracy based on question type for free version of ChatGPT

| Analysis by Question Type: Free Version | Total Questions per type | Right by category | Percentage |
|---|---:|---:|---:|
| Multiple Choice | 39 | 22 | 56.41% |
| Fill in the Blank | 5 | 1 | 20.00% |
| Matching | 1 | 1 | 100.00% |
| Ordering | 0 | 0 | Undefined |
| All that apply | 2 | 0 | 0.00% |

Conclusion

An objective of this paper is to expand upon the set of literature that has attempted to analyze the abilities of ChatGPT to be successful in various disciplines. To that end we have focused on putting two models of ChatGPT, the freely available version (GPT-3.5) and the subscription based GPT-4, through the mechanical engineering FE exam. The NCEES does not provide the information needed to determine what it takes to pass the FE exam, but due to the limitation for text only inputs and the correct response rates seen in our data we feel it is unlikely either model would pass at this point. However, it is clear from our data that the GPT-4 model greatly out performs its free counterpart. Additionally, our method of input mimics how an undergraduate student might approach using ChatGPT to assist with problem solving which provides information regarding both the potential use cases, as well as pitfalls, of ChatGPT from both student and instructor perspectives.

We found that, compared to its free counterpart, GPT-4 was a superior tool for engaging with engineering questions. It consistently provided more detail in its answers and was 50% more likely to be accurate based on our small dataset. We saw that GPT-4 performed well in nearly every topic it was given and could handle a variety of question types. Additionally, GPT-4 was more likely than the free version to avoid making invalid assumptions and attempting to solve problems without all necessary information. However, both models were capable of providing detailed misinformation. We can confirm what was seen in the literature where both models made errors in arithmetic and units, but often presented those errors with confidence as correct answers. This makes it clear that anyone using ChatGPT to solve engineering problems must be willing to double check its work. Fortunately ChatGPT does allow for conversations, so if an error in assumption or equation is found a follow up question can be asked, but as seen in our work and other literature follow up and prompt engineering do not guarantee a correct response. Another pitfall of ChatGPT at the time of writing this article is that it does not have a mechanism for visual input. From an instructor's perspective this may be treated as an advantage. Instructors concerned with students using ChatGPT as a tool to cheat on homework or exams may be able to combat that by incorporating questions that make use of plots, charts, or other visual schematics. These findings support what is available in the literature surrounding the use of ChatGPT in other STEM disciplines.

Based on our research and the conclusions drawn from the literature we consulted for our review, we have identified possible applications of ChatGPT in the engineering classroom. As an LLM, ChatGPT's ability to pull information from a variety of sources and identify patterns may make it a useful assistant when developing lesson plans and assignments. Educators, as experts capable of verifying the accuracy of resources in their given domains, can use ChatGPT to develop materials for their courses and may be able to use the tool to develop differentiated learning materials for students of varying skill levels. Using ChatGPT to facilitate a flipped classroom may also present an opportunity to embrace the technology rather than discouraging its use - engineering faculty may also be able to leverage the tool's inconsistency and inaccuracy to spark discussion among students and sharpen their critical thinking and problem solving skills

(Leffer, 2023). ChatGPT's ability to provide explanations and its broad training data can also make it a useful tool for students seeking background information on general topics and definitions for specific terms with the caveat that students should corroborate its answers with other sources. This may add to students' learning experiences by offering them an outlet to ask questions and receive personalized feedback without the fear (imagined or real) of judgment associated with asking questions in class or during office hours. Last, as Meneske (2023) suggests, ChatGPT may also be used to create virtual learning environments and simulations of real-life engineering situations that can help students visualize and apply engineering concepts in novel ways.

We have also identified a possible approach educators can take to mitigate negative impacts associated with ChatGPT's use, such as cheating. As shown in our literature review and in ChatGPT and GPT-4's difficulty in answering a multistep open ended questions from the FE exam and junior and senior level course exams, ChatGPT does not perform well when tasked with applying knowledge to unique and novel situations, producing complex analyses that require the application and interpretation of knowledge, and answering context-dependent questions. Rather than employing punitive approaches to prevent students from using ChatGPT, educators can use the disruptive impact of this tool to reimagine their assignments and create questions and assessments that are difficult for LLMs to process (Fergus et al., 2023). Encouraging students to be transparent in how they answer questions and focusing the process of problem solving rather than arriving at a specific numeric answer may also be a useful approach in building trust (i.e., rather than assuming malintent) and increasing students' understanding of how to develop strong problem solving skills.

In professional settings, ChatGPT may be a useful assistant for professionals seeking to refresh their knowledge of specific topics and as a supplement to searches for information. However, caution should be exercised when considering ChatGPT as a mathematical tool due to its tendency to confidently produce incorrect answers and its functionality as an LLM that predicts responses rather than performing calculations. Therefore, professionals that intend to use ChatGPT in the workplace may want to avoid using the tool to make automated decisions and include a verification step in any workflow that utilizes the ChatGPT.  In both professional and educational contexts, greater transparency around the design, training data, and development of ChatGPT would aid users in understanding the tools' biases and limitations, allowing for more effective and informed use.

Though a big question still left unanswered is how critical students and researchers are of the information they receive from ChatGPT and similar programs. To address this final question the authors have started a second research study focused on examining how engineering students engage with ChatGPT. The study will be conducted in two parts. The first will be a survey used to gather information on how students are engaging with ChatGPT. The second part will consist of observing students' use of ChatGPT to answer preset questions and then interview those students about their choices and confidence in the results.

While ChatGPT has the potential to be a capable tool for users in many disciplines, including engineering, it is clear that due to its inconsistency and tendency to confidently produce incorrect answers it is best suited for users with expert knowledge. The introduction of the calculator raised concerns among math educators around whether or not students would be able to learn the mechanisms behind their calculations if a calculator was permitted. Similar concerns were raised with the advent of the PC, increased access to the internet, and, now, with access to ChatGPT. One of the key distinctions between these tools is their reliability. Calculators do not commonly produce incorrect calculations and educators have found ways to incorporate their use in the classroom without compromising students' ability to solve problems. Access to the internet was seen as another disruptive tool and, while the reliability of information on the internet is a constant cause for caution, the use of PCs and internet has become a commonplace tool in the classroom. ChatGPT may present the next disruptive technology in science and engineering, but it may be premature to claim the current models as such. Much like the calculator and the PC before it, ChatGPT may be poised as the next opportunity for educators to reevaluate their methods and can make the case for increased critical thinking and information literacy skills in all disciplines. If ChatGPT is here to stay, further exploration and examination of how ChatGPT can or should be incorporated into STEM education and professions is vital.

References

Abramski, K., Citraro, S., Lombardi, L., Rossetti, G., & Stella, M. (2023). *Cognitive network science reveals bias in GPT-3, ChatGPT, and GPT-4 mirroring math anxiety in high-school students* (arXiv:2305.18320). arXiv. https://doi.org/10.48550/arXiv.2305.18320

Azaria, A., Azoulay, R., & Reches, S. (2023). *ChatGPT is a Remarkable Tool -- For Experts*. https://doi.org/10.48550/ARXIV.2306.03102

Bordt, S., & von Luxburg, U. (2023). *ChatGPT Participates in a Computer Science Exam*. https://doi.org/10.48550/ARXIV.2303.09461

Borji, A. (2023). *A Categorical Archive of ChatGPT Failures* (arXiv:2302.03494). arXiv. http://arxiv.org/abs/2302.03494

Chen, L., Zaharia, M., & Zou, J. (2023). *How is ChatGPT's behavior changing over time?* https://doi.org/10.48550/ARXIV.2307.09009

Delua, J. (2021, March 12). Supervised vs. unsupervised learning: What's the difference? *IBM Blog.* https://www.ibm.com/blog/supervised-vs-unsupervised-learning/

Dennean, K., Gantori, S., Limas, D. K., Pu, A., Gilligan, R. (2023, February 22). *Let's chat about ChatGPT.* UBS. https://secure.ubs.com/public/api/v2/investment-content/documents/XILxY9V9P5RazGpDA1Cr_Q?apikey=Y8VdAx8vhk1P9YXDlEOo2Eoco1fqKwDk&Accept-Language=de-CH

Deshpande, S., & Szefer, J. (2023). *Analyzing ChatGPT's Aptitude in an Introductory Computer Engineering Course* (arXiv:2304.06122). arXiv. https://doi.org/10.48550/arXiv.2304.06122

Elder, C., Pozek, G., Horine, S., Tripaldelli, A., & Butka, B. (2023). Can Artificial Intelligence Pass a Sophomore Level Digital Design Laboratory? *SoutheastCon 2023*, 861–868. https://doi.org/10.1109/SoutheastCon51012.2023.10115116

Fergus, S., Botha, M., & Ostovar, M. (2023). Evaluating Academic Answers Generated Using ChatGPT. *Journal of Chemical Education*, *100*(4), 1672–1675. https://doi.org/10.1021/acs.jchemed.3c00087

Fijačko, N., Gosak, L., Štiglic, G., Picard, C. T., & John Douma, M. (2023). Can ChatGPT pass the life support exams without entering the American heart association course? *Resuscitation*, *185*, 109732. https://doi.org/10.1016/j.resuscitation.2023.109732

Frieder, S., Pinchetti, L., Griffiths, R.-R., Salvatori, T., Lukasiewicz, T., Petersen, P. C., Chevalier, A., & Berner, J. (2023). *Mathematical Capabilities of ChatGPT*. https://doi.org/10.48550/ARXIV.2301.13867

Geng, C., Zhang, Y., Pientka, B., & Si, X. (2023). *Can ChatGPT Pass An Introductory Level Functional Language Programming Course?* https://doi.org/10.48550/ARXIV.2305.02230

Giannos, P., & Delardas, O. (2023). Performance of ChatGPT on UK Standardized Admission Tests: Insights From the BMAT, TMUA, LNAT, and TSA Examinations. *JMIR Medical Education*, *9*(1), e47737. https://doi.org/10.2196/47737

Gilson, A., Safranek, C. W., Huang, T., Socrates, V., Chi, L., Taylor, R. A., & Chartash, D. (2023). How Does ChatGPT Perform on the United States Medical Licensing Examination? The Implications of Large Language Models for Medical Education and Knowledge Assessment. *JMIR Medical Education*, *9*, e45312.

https://doi.org/10.2196/45312

Grose, T. K. (2023). Disruptive influence: Generative AI offers both a powerful temptation to cheat and a potent opportunity to learn. *American Society for Engineering Education 32*(3): 16-17.

Hartman-Sigall, J. (2023, January 25). University declines to ban ChatGPT, releases faculty guidance for its usage. *The Daily Princetonian.*

https://www.dailyprincetonian.com/article/2023/01/university-declines-ban-chatgpt-releases-faculty-guidance-for-usage

Hashana, A. M. J., Brundha, P., Ahamed Ayoobkhan, M. U., & S, F. (2023). Deep Learning in ChatGPT - A Survey. *2023 7th International Conference on Trends in Electronics and Informatics (ICOEI)*, 1001–1005.

https://doi.org/10.1109/ICOEI56765.2023.10125852

Humar, P., Asaad, M., Bengur, F. B., & Nguyen, V. (2023). ChatGPT is Equivalent to First Year Plastic Surgery Residents: Evaluation of ChatGPT on the Plastic Surgery In-Service Exam. *Aesthetic Surgery Journal*, sjad130.

https://doi.org/10.1093/asj/sjad130

Johri, A., Katz, A. S., Qadir, J., & Hingle, A. (2023). Generative artificial intelligence and engineering education. *Journal of Engineering Education*, *112:* 572-577.

https://doi.org/10.1002/jee.20537

Kasai, J., Kasai, Y., Sakaguchi, K., Yamada, Y., & Radev, D. (2023). *Evaluating GPT-4 and ChatGPT on Japanese Medical Licensing Examinations.*

https://doi.org/10.48550/ARXIV.2303.18027

Kortemeyer, G. (2023). *Could an Artificial-Intelligence agent pass an introductory physics course?* https://doi.org/10.48550/ARXIV.2301.12127

Koubaa, A., Qureshi, B., Ammar, A., Khan, Z., Boulila, W., & Ghouti, L. (2023). *Humans are Still Better than ChatGPT: Case of the IEEEXtreme Competition*. https://doi.org/10.48550/ARXIV.2305.06934

Kung, T. H., Cheatham, M., Medenilla, A., Sillos, C., De Leon, L., Elepaño, C., ... & Tseng, V. (2023). Performance of ChatGPT on USMLE: Potential for AI-assisted medical education using large language models. *PLoS digital health*, *2*(2), e0000198. https://doi.org/10.1371/journal.pdig.0000198

Laskar, M. T. R., Bari, M. S., Rahman, M., Bhuiyan, M. A. H., Joty, S., & Huang, J. X. (2023). *A Systematic Study and Comprehensive Evaluation of ChatGPT on Benchmark Datasets*. https://doi.org/10.48550/ARXIV.2305.18486

Leffer, L. (2023, August 25). *ChatGPT can get good grades. What should educators do about it?* Scientific American. Retrieved from https://www.scientificamerican.com/article/chatgpt-can-get-good-grades-what-should-educators-do-about-it/

Maloney, E. A., Schaeffer, M. W., & Beilock, S. L. (2013). Mathematics anxiety and stereotype threat: shared mechanisms, negative consequences and promising interventions. *Research in Mathematics Education*, *15*(2), 115–128. https://doi.org/10.1080/14794802.2013.797744

Menekse, M. (2023), Envisioning the future of learning and teaching engineering in the artificial intelligence era: Opportunities and challenges. *Journal of Engineering Education, 112:* 578-582. https://doi.org/10.1002/jee.20539

Natalie (2023). *ChatGPT - release notes.*

https://help.openai.com/en/articles/6825453-chatgpt-release-notes

NCEES (n.d.). *FE Exam*. https://ncees.org/exams/fe-exam/

Nield, D. (2023, April 30). How ChatGPT and other LLMs work - and where they could go next. *WIRED.*

https://www.wired.com/story/how-chatgpt-works-large-language-model/

Ogundare, O., Madasu, S., & Wiggins, N. (2023). *Industrial Engineering with Large Language Models: A case study of ChatGPT's performance on Oil & Gas problems*.

https://doi.org/10.48550/ARXIV.2304.14354

OpenAI. (2022, November 30). *Introducing ChatGPT.* https://openai.com/blog/chatgpt

OpenAI. (2023a). *GPT-4 Technical Report* (arXiv:2303.08774). arXiv.

http://arxiv.org/abs/2303.08774

OpenAI. (2023b, February 1). *Introducing ChatGPT plus.*

https://openai.com/blog/chatgpt-plus

OpenAI. (2023c, March 14). *GPT-4*. https://openai.com/research/gpt-4

OpenAI. (2023d, March 18). *ChatGPT plugins.* https://openai.com/blog/chatgpt-plugins

OpenAI. (2023e, May 18). *Introducing the ChatGPT app for iOS.*

https://openai.com/blog/introducing-the-chatgpt-app-for-ios

Plevris, V., Papazafeiropoulos, G., & Rios, A. J. (2023). *Chatbots put to the test in math and logic problems: A preliminary comparison and assessment of ChatGPT-3.5, ChatGPT-4, and Google Bard*. https://doi.org/10.48550/ARXIV.2305.18618

Pursnani, V., Sermet, Y., & Demir, I. (2023). *Performance of ChatGPT on the US Fundamentals of Engineering Exam: Comprehensive Assessment of Proficiency and*

*Potential Implications for Professional Environmental Engineering Practice*.

https://doi.org/10.48550/ARXIV.2304.12198

Radford, A., Narasimhan, K., Salimans, T., & Sutskever, I. (2018). Improving language understanding by generative pre-training.

https://openai.com/research/language-unsupervised

Sallam, M., Salim, N. A., Al-Tammemi, A. B., Barakat, M., Fayyad, D., Hallit, S., Harapan, H., Hallit, R., & Mahafzah, A. (2023). Chatgpt output regarding compulsory vaccination and covid-19 vaccine conspiracy: A descriptive study at the outset of a paradigm shift in online search for information. *Cureus*, *15*(2), e35029.

https://doi.org/10.7759/cureus.35029

Satariano, A. (2023, March 31). ChatGPT is banned in Italy over privacy concerns. *The New York Times*.

https://www.nytimes.com/2023/03/31/technology/chatgpt-italy-ban.html?searchResultPosition=2

Shakarian, P., Koyyalamudi, A., Ngu, N., & Mareedu, L. (2023). *An Independent Evaluation of ChatGPT on Mathematical Word Problems (MWP)*.

https://doi.org/10.48550/ARXIV.2302.13814

Stoke-Walker, C. (2023, January 18). ChatGPT listed as author on research papers: many scientists disapprove. *Nature*.

https://www.nature.com/articles/d41586-023-00107-z

van Dis, E. A. M., Bollen, J., Zuidema, W., van Rooij, R., & Bockting, C. L. (2023). ChatGPT: five priorities for research. *Nature*, *614*(7947), 224–226.

https://doi.org/10.1038/d41586-023-00288-7

Wu, T., He, S., Liu, J., Sun, S., Liu, K., Han, Q.-L., & Tang, Y. (2023). A Brief Overview of ChatGPT: The History, Status Quo and Potential Future Development. *IEEE/CAA Journal of Automatica Sinica*, *10*(5), 1122–1136. https://doi.org/10.1109/JAS.2023.123618

Xuan-Quy, D., Ngoc-Bich, L., Xuan-Dung, P., Bac-Bien, N., & The-Duy, V. (2023). *Evaluation of ChatGPT and Microsoft Bing AI Chat Performances on Physics Exams of Vietnamese National High School Graduation Examination*. https://doi.org/10.48550/ARXIV.2306.0453

Yuan, Z., Yuan, H., Tan, C., Wang, W., & Huang, S. (2023). *How well do Large Language Models perform in Arithmetic tasks?* https://doi.org/10.48550/ARXIV.2304.02015

Appendix A - Verbatim Responses from ChatGPT

Table A. Questions and responses given by different models of ChatGPT for sample set questions.

| Category | Question Text | GPT-4 Response | Freely available model |
|---|---|---|---|
| Both Correct | The Pole locations of a transfer function on the complex plane provides us information about<br><br>A. Naturally Frequency<br>B. Damping Ratio<br>C. System Stability<br>D. All of the above | D. All of the above<br><br>The pole locations of a transfer function on the complex plane provide information about all three of the mentioned properties: natural frequency, damping ratio, and system stability.<br><br>- Natural frequency: The distance from the origin of | D. All of the above.<br><br>The pole locations of a transfer function on the complex plane provide information about the characteristics of the system. Specifically:<br><br>1. Natural Frequency: The distance of the poles from the origin on the complex plane relates to the system's natural |
| | | the complex plane to the pole represents the natural frequency of the system. Natural frequency refers to the frequency at which a system naturally oscillates in the absence of any driving forces.<br><br>- Damping ratio: The angle of the line from the origin to the pole in the complex plane, typically measured relative to the negative real axis, represents the damping ratio of the system. The damping ratio indicates how quickly the system returns to equilibrium after being disturbed.<br><br>- System stability: If all the poles of the system lie in the left half of the complex plane, the system is stable. If any poles lie in the right half, the system is unstable. If poles lie on the imaginary axis (i.e., real part is zero), the system is marginally stable.<br><br>Thus, the pole locations are quite informative about various properties of the system represented by the transfer function. | frequency. The further away the poles are from the origin, the higher the natural frequency of the system.<br><br>2. Damping Ratio: The relative placement of the poles on the complex plane determines the damping ratio of the system. The angle and separation between the poles can indicate whether the system is underdamped, critically damped, or overdamped.<br><br>3. System Stability: The stability of a system is determined by the pole locations. If all the poles of a system lie in the left-half of the complex plane (i.e., have negative real parts), the system is stable. If any poles are in the right-half of the complex plane (i.e., have positive real parts), the system is unstable.<br><br>Therefore, the pole locations of a transfer function on the complex plane provide information about the system's natural frequency, damping ratio, and stability. |
| Only 4.0 Correct | A white dwarf star is estimated to have a temperature of 100,000 Kelvin. At what wavelength would the spectral distribution have its maximum value. | The peak wavelength of a black body, such as a star, can be determined by Wien's displacement law:<br><br>$\lambda\_max * T = b$<br><br>where: | To determine the wavelength at which the spectral distribution of a white dwarf star has its maximum value, we can use Wien's displacement law. According to this law, the wavelength of maximum intensity ($\lambda\_max$) is |

|  |  | A. 0.029\mu m<br>B. 0.29\mu m<br>C. 2.9\mu m<br>D. 29\mu m | $\lambda_{max}$ is the wavelength at which the distribution peaks, T is the absolute temperature of the black body, and<br>b is a constant known as Wien's displacement constant, approximately equal to 2.898 x 10^-3 m.K<br><br>We need to solve for $\lambda_{max}$, so we can rearrange the formula to:<br><br>$\lambda_{max}$ = b / T<br><br>Given that T = 100,000 K and b = 2.898 x 10^-3 m.K, we substitute these values into the equation:<br><br>$\lambda_{max}$ = 2.898 x 10^-3 m.K / 100,000 K<br>$\lambda_{max}$ = 2.898 x 10^-8 m<br><br>This is equivalent to 0.029 x 10^-6 m, or 0.029 micrometers (μm).<br>Therefore, option A is the correct answer. | inversely proportional to the temperature (T) of the object. Mathematically, the relationship is expressed as:<br><br>$\lambda_{max}$ = b / T<br><br>where $\lambda_{max}$ is the wavelength at maximum intensity, b is Wien's displacement constant (approximately equal to 2.898 × 10^-3 m·K), and T is the temperature in Kelvin.<br><br>In this case, the temperature of the white dwarf star is given as 100,000 Kelvin. Plugging the values into the equation:<br><br>$\lambda_{max}$ = (2.898 × 10^-3 m·K) / (100,000 K)<br><br>Calculating this expression, we find:<br><br>$\lambda_{max}$ ≈ 2.898 × 10^-8 m ≈ 29 × 10^-9 m<br><br>Converting meters to micrometers (μm), we get:<br><br>$\lambda_{max}$ ≈ 29 μm<br><br>Therefore, the spectral distribution of the white dwarf star would have its maximum value at a wavelength of approximately 29 μm. Thus, the correct answer is option D. |
| Only Free Correct | A sphere of material is found in a flow with h=200W/m^2K has a radius of 1 | The Biot number (Bi) is a dimensionless quantity used in heat transfer calculations. It is defined as: | The Biot number (Bi) is a dimensionless number that characterizes the relative importance of internal thermal resistance within a solid |

| | millimeter if the sphere has a thermal conductivity of k=20 W/mK what is the Biot number of this system and does it satisfy the Lumped Capacitance criteria?<br><br>A. 0.0033, No to Lumped Capacitance<br>B. 0.0033, Yes to Lumped Capacitance<br>C. 0.33, No to Lumped Capacitance<br>D. 0.33, Yes to Lumped Capacitance | Bi = hLc/k<br><br>where:<br>h is the heat transfer coefficient,<br>Lc is the characteristic length (in this case, the radius of the sphere),<br>k is the thermal conductivity.<br><br>The Lumped Capacitance method is only valid if the Biot number is less than 0.1. This implies that heat conduction within the object is much faster than heat convection away from the object, so the temperature within the object can be considered uniform.<br><br>In this case, we can calculate the Biot number using the provided values:<br><br>h = 200 W/m²K<br>Lc = 1 millimeter = 0.001 meter<br>k = 20 W/mK<br><br>So,<br><br>Bi = h*Lc/k = (200 W/m²K * 0.001 m) / (20 W/mK) = 0.01.<br><br>The Biot number is less than 0.1, therefore the system should satisfy the Lumped Capacitance criteria.<br><br>There seems to be a misunderstanding as none of the choices you provided match with the calculated Biot number and the answer | compared to the thermal resistance at the surface. It is defined as the ratio of the internal conduction resistance to the external convection resistance.<br><br>The Biot number (Bi) can be calculated using the following formula:<br><br>Bi = h * r / k,<br><br>where h is the convective heat transfer coefficient, r is the radius of the sphere, and k is the thermal conductivity of the material.<br><br>In this case, the convective heat transfer coefficient (h) is given as 200 W/m^2K, the radius of the sphere (r) is 1 millimeter (which is 0.001 meters), and the thermal conductivity of the material (k) is 20 W/mK.<br><br>Plugging in the values into the formula, we have:<br><br>Bi = (200 W/m^2K) * (0.001 m) / (20 W/mK) = 0.01.<br><br>So, the Biot number for this system is 0.01.<br><br>The Lumped Capacitance criteria can be satisfied when the Biot number is less than 0.1. In this case, since the Biot number is 0.01, which is less than 0.1, the system satisfies the Lumped Capacitance criteria. |

| | | to the Lumped Capacitance criteria. However, based on the calculations, the Biot number would be 0.01 and yes, the system satisfies the Lumped Capacitance criteria. | Therefore, the correct answer is B. 0.0033, Yes to Lumped Capacitance. |
|---|---|---|---|
| Both Incorrect | A surface with temperature of 300^\circ C has emissivity of \epsilon=0.8. It is contained in a large surrounding with temperature of 30^\circ C. Air at 30^\circ C blows over the plate. At what coefficient of heat transfer will the convective heat flux match the radiative heat flux?<br><br>A. 1.4 W/m^2K<br>B. 10.2 W/m^2K<br>C. 16.7 W/m^2K<br>D. 32.2 W/m^2K | The radiative heat transfer between a surface and its surroundings is given by the Stefan-Boltzmann law, which states:<br><br>q_rad = εσ($T_1^4$ - $T_2^4$)<br><br>where:<br>q_rad = radiative heat flux,<br>ε = emissivity of the surface,<br>σ = Stefan-Boltzmann constant (~5.67 × $10^{-8}$ W/m²K⁴),<br>$T_1$ = temperature of the surface (in Kelvin),<br>$T_2$ = temperature of the surroundings (in Kelvin).<br><br>The convective heat transfer between a surface and a fluid is given by:<br><br>q_conv = h($T_1$ - $T_2$)<br><br>where:<br>q_conv = convective heat flux,<br>h = coefficient of heat transfer,<br>$T_1$ = temperature of the surface (in Celsius),<br>$T_2$ = temperature of the fluid (in Celsius).<br><br>Since we are given that the radiative heat flux matches the convective heat flux, we | To determine the coefficient of heat transfer at which the convective heat flux matches the radiative heat flux, we need to calculate the convective and radiative heat fluxes separately and equate them.<br><br>The convective heat transfer rate can be calculated using Newton's law of cooling, which states:<br><br>Q_conv = h × A × (T_surface - T_surrounding)<br><br>Where:<br>Q_conv is the convective heat transfer rate,<br>h is the coefficient of heat transfer,<br>A is the surface area,<br>T_surface is the temperature of the surface, and<br>T_surrounding is the temperature of the surrounding air.<br><br>The radiative heat transfer rate can be calculated using the Stefan-Boltzmann law, which states:<br><br>Q_rad = ε × σ × A × (T_surface^4 - T_surrounding^4) |

| | | can set the two equations equal to each other: $\varepsilon\sigma(T_1^4 - T_2^4) = h(T_1 - T_2)$ Given the values: $\varepsilon = 0.8$, $\sigma = 5.67 \times 10^{-8}$ W/m²K⁴, $T_1 = 300°C = 573.15$ K (converted from Celsius to Kelvin), $T_2 = 30°C = 303.15$ K (converted from Celsius to Kelvin). We can solve for h: $h = \varepsilon\sigma(T_1^4 - T_2^4) / (T_1 - T_2)$ Substitute the values into the equation: $h = 0.8 \times 5.67 \times 10^{-8} (573.15^4 - 303.15^4) / (573.15 - 303.15)$ Solving this gives approximately 10.2 W/m²K. Therefore, option B is correct. | Where: $Q\_rad$ is the radiative heat transfer rate, $\varepsilon$ is the emissivity of the surface, $\sigma$ is the Stefan-Boltzmann constant ($5.67 \times 10^{-8}$ W/m^2K^4), A is the surface area, $T\_surface$ is the temperature of the surface, and $T\_surrounding$ is the temperature of the surrounding air. We can set these two equations equal to each other and solve for the coefficient of heat transfer (h): $h \times A \times (T\_surface - T\_surrounding) = \varepsilon \times \sigma \times A \times (T\_surface^4 - T\_surrounding^4)$ Simplifying the equation: $h = \varepsilon \times \sigma \times (T\_surface^4 - T\_surrounding^4) / (T\_surface - T\_surrounding)$ Plugging in the given values: $T\_surface = 300°C = 573$ K $T\_surrounding = 30°C = 303$ K $\varepsilon = 0.8$ $\sigma = 5.67 \times 10^{-8}$ W/m^2K^4 $h = 0.8 \times 5.67 \times 10^{-8} \times (573^4 - 303^4) / (573 - 303)$ $h \approx 1.36$ W/m^2K The coefficient of heat transfer at which the convective heat flux matches the radiative heat flux is approximately 1.36 W/m^2K. |

| | | | Since none of the answer choices match exactly, we can round the value to the nearest option. In this case, the closest option is A. 1.4 W/m^2K. |
|---|---|---|---|
| Free Response (GPT-4 Correct, Free Version Incorrect) | Air at 27◦ C and 1 atm flows over a flat plate at a speed of 2 m/s. The plate is 40 cm long, 1 cm wide and is heated over its entire length to a temperature of 60◦ C.<br>1. What is the film temperature?<br>2. Determine the values of density, kinematic viscosity, thermal conductivity and Prandtl number required for this problem.<br>3. Compute the Reynolds number using a relevant characteristic length.<br>4. What is an appropriate Nusselt number correlation for this problem?<br>5. Using the Nusselt number correlation that you identified above, determine the convective heat transfer coefficient.<br>6. Calculate the heat loss from the entire plate. | This seems to be a classic heat transfer problem involving flow over a flat plate. Here are the steps to solve this:<br><br>1. **Film Temperature:**<br><br>The film temperature (T_f) is usually taken as the average of the free stream temperature (T_inf) and the surface temperature (T_s). The given temperatures need to be converted from Celsius to Kelvin.<br><br>T_inf = 27°C + 273.15 = 300.15 K<br>T_s = 60°C + 273.15 = 333.15 K<br><br>T_f = (T_s + T_inf)/2 = (333.15K + 300.15K) / 2 = 316.65 K<br><br>2. **Air Properties at the Film Temperature:**<br><br>At the film temperature, the values of density (ρ), kinematic viscosity (ν), thermal conductivity (k), and the Prandtl number (Pr) for air can be obtained from standard air property tables or correlations. As of my knowledge cutoff in 2021, typical values for air at ~316 | To solve the given problem, we will go step by step:<br><br>1. Film temperature (T_f):<br>The film temperature is the average temperature between the surface temperature of the plate and the temperature of the fluid flow. In this case, the surface temperature of the plate is 60°C, and the air temperature is 27°C. Therefore, the film temperature is:<br>T_f = (T_plate + T_air) / 2<br>= (60 + 27) / 2<br>= 87 / 2<br>= 43.5°C<br><br>2. Required properties:<br>To determine the values of density (ρ), kinematic viscosity (ν), thermal conductivity (k), and Prandtl number (Pr), we can use the properties of air at the film temperature (T_f). You can look up these values in tables or use correlations. For simplicity, I will provide approximate values at 40°C:<br><br>Density (ρ) of air at 40°C: ρ ≈ 1.164 kg/m³<br>Kinematic viscosity (ν) of air at 40°C: ν ≈ 1.348 × 10^(-5) m²/s<br>Thermal conductivity (k) of air at 40°C: k ≈ 0.0263 W/(m·K)<br>Prandtl number (Pr) of air at 40°C: Pr ≈ 0.712 |

| | | K (43 °C) are:<br><br>ρ ~ 1.127 kg/m^3,<br>v ~ 20.92 x 10^-6 m^2/s,<br>k ~ 0.028 W/m·K,<br>Pr ~ 0.7.<br><br>Please verify these values with an updated air properties table or correlation.<br><br>3. **Reynolds Number:**<br><br>The Reynolds number (Re) is given by Re = ρUx/μ = Ux/v, where U is the free stream velocity, x is the characteristic length, and μ is the dynamic viscosity. In this case, the characteristic length would be the length of the plate, which is 0.4 m. The free stream velocity is given as 2 m/s.<br><br>Re = (2 m/s * 0.4 m) / (20.92 x 10^-6 m^2/s) = 38168<br><br>4. **Nusselt Number Correlation:**<br><br>Since the Reynolds number is less than 5 x 10^5, we are dealing with a laminar flow over a flat plate, and we can use the correlation for Nusselt number for a laminar boundary layer over a flat plate, which is Nu = 0.664 Re^0.5 Pr^(1/3), assuming a Prandtl number of around 0.7 for air.<br><br>5. **Convective Heat Transfer Coefficient (h):** | 3. Reynolds number (Re): The Reynolds number is a dimensionless quantity that relates the inertial forces to the viscous forces and determines the flow regime. It can be calculated using a relevant characteristic length (L), which in this case is the length of the plate.<br><br>Reynolds number (Re) = (ρ * U * L) / v<br>where U is the velocity of the air flow and L is the length of the plate.<br><br>Re = (ρ * U * L) / v<br>= (1.164 * 2 * 0.4) / 1.348 × 10^(-5)<br>≈ 6.83 × 10^5<br><br>4. Nusselt number correlation: The appropriate Nusselt number correlation depends on the flow regime. Since the problem does not specify the flow regime, we can assume it to be laminar or turbulent. For laminar flow over a flat plate, the Nusselt number correlation can be given by:<br><br>Nu = 0.332 * Re^(0.5) * Pr^(1/3)<br><br>For turbulent flow, the correlation is more complex and depends on additional factors.<br><br>5. Convective heat transfer coefficient (h): The convective heat transfer coefficient can be determined |

| | | The Nusselt number is defined as Nu = h*x/k. From this, we can solve for h:<br><br>Nu = 0.664 * Re^0.5 * Pr^(1/3) = 0.664 * (38168)^0.5 * 0.7^(1/3) = 122.79<br><br>h = Nu * k / x = 122.79 * 0.028 W/m·K / 0.4 m = 8.6 W/m^2·K<br><br>6. **Heat Loss from the Plate:**<br><br>The heat loss from the entire plate can be calculated using the equation for convection: q = h*A*ΔT, where A is the area of the plate and ΔT is the temperature difference between the surface and the free stream.<br><br>A = Length * Width = 0.4 m * 0.01 m = 0.004 m^2<br><br>ΔT = T_s - T_inf = 333.15 K - 300.15 K = 33 K<br><br>q = h * A * ΔT = 8.6 W/m^2·K * 0.004 m^2 * 33 K = 1.14 W.<br><br>Please note:<br><br>All these calculations assume ideal conditions and neglect influences such as boundary layer development, surface roughness, and so on. They | using the Nusselt number (Nu) correlation obtained above and the thermal conductivity (k) of air.<br><br>Nu = h * L / k<br>h = Nu * k / L<br><br>Assuming laminar flow, we can use the Nusselt number correlation:<br><br>Nu = 0.332 * Re^(0.5) * Pr^(1/3)<br>h = (0.332 * Re^(0.5) * Pr^(1/3) * k) / L<br><br>Substituting the known values:<br>h = (0.332 * (6.83 × 10^5)^(0.5) * (0.712)^(1/3) * 0.0263) / 0.4<br>≈ 145.28 W/(m²·K)<br><br>6. Heat loss from the entire plate:<br>The heat loss can be calculated using the convective heat transfer coefficient (h), the surface area of the<br><br>plate (A), and the temperature difference between the plate and the fluid flow (ΔT).<br><br>Q = h * A * ΔT<br><br>The surface area of the plate is given by:<br>A = length * width<br><br>Substituting the known values:<br>A = 0.4 * 0.01 = 0.004 m²<br>(since width is given in cm)<br>ΔT = T_plate - T_air = 60 - 27 = 33°C |

| | | should therefore be used for estimation purposes only. Also, the properties of air can vary, so it's important to use an up-to-date source when solving these types of problems. | Q = 145.28 * 0.004 * 33 ≈ 191.37 W

Therefore, the heat loss from the entire plate is approximately 191.37 Watts. |
|---|---|---|---|